# OSPF WEIGHT SETTING OPTIMIZATION FOR SINGLE LINK FAILURES

Mohammed H. Sqalli, Sadiq M. Sait, and Syed Asadullah

Computer Engineering Department
King Fahd University of Petroleum & Minerals
Dhahran 31261, Saudi Arabia
{sqalli,sadiq,sasad}@kfupm.edu.sa

*ABSTRACT*

*In operational networks, nodes are connected via multiple links for load sharing and redundancy. This is done to make sure that a failure of a link does not disconnect or isolate some parts of the network. However, link failures have an effect on routing, as the routers find alternate paths for the traffic originally flowing through the link which has failed. This effect is severe in case of failure of a critical link in the network, such as backbone links or the links carrying higher traffic loads. When routing is done using the Open Shortest Path First (OSPF) routing protocol, the original weight selection for the normal state topology may not be as efficient for the failure state. In this paper, we investigate the single link failure issue with an objective to find a weight setting which results in efficient routing in normal and failure states. We engineer Tabu Search Iterative heuristic using two different implementation strategies to solve the OSPF weight setting problem for link failure scenarios. We evaluate these heuristics and show through experimental results that both heuristics efficiently handle weight setting for the failure state. A comparison of both strategies is also presented.*

## 1. INTRODUCTION

OSPF is an intra-domain routing protocol that uses link weights to make routing decisions and compute the shortest paths. Different weight assignment strategies have been discussed in the literature [11] including the Unit OSPF, Inverse Capacity OSPF, Random OSPF etc. A better selection of the OSPF link weights can lead to efficient network utilization [1, 2]. Iterative heuristics have been extensively used [10, 3, 4] and implemented using different strategies to achieve this goal.

Ericsson et al. [4] proposed a Genetic Algorithm and used the set of test problems considered in [11]. A hybrid GA was also proposed by them [5] which makes use of the dynamic shortest path algorithm to recompute shortest paths after the modification of link weights. Sridharan et al. [6] developed another heuristic for a slightly different version of the problem, in which the flow is split among a subset of the outgoing links on the shortest paths to the destination IP address.

In this work, we have used a Tabu Search (TS) algorithm [7] to solve the OSPFWS problem. Tabu search is an iterative heuristic that has been applied for solving a range of combinatorial optimization problems in different fields. The detailed description and related references for tabu search can be found in [7].

However, all strategies work on the assumption that the topology is fixed and there are no failures in the network. A network may experience a link failure resulting in a change in topology due to the loss of a link, when the network state changes (Failure State) due to link failure, the routing paths are also not the same as in the original state (Normal State). The optimized weights for the original topology and demand may no longer be good enough for the new topology with the failed link. The absence of the failed link causes the traffic which was originally flowing through this link to flow through other available links. The fact that the network was not optimized for these flows can result in an inefficient mapping of traffic on to available links. This may also cause congestion in some parts of the network, especially in the case of higher demands.

One solution to this issue is to apply a new set of OSPF weights to links which optimize the new topology (Failure State). However, it is cumbersome to change the weights on each link in the entire topology and also not very practical in case of larger networks. One would suppose that once the set of OSPF weights have been fixed, the operator would not want to change these weights in order to adapt for such changes in the state of the network. Hence, it is required to adapt the original heuristic to optimize link weights taking into consideration single link failure scenarios. In other words it is required to find a set of weights that work for both the normal and failed state of the network without considerable degradation in performance in both states.

Link failure scenarios require dealing with two states of a network. The first state where all links are functional is denoted as Normal state and the other state where a link has failed is denoted as Failure state. In this paper, which is an extension of Sqalli et al. [8], two different strategies are devised and implemented to address this issue. The first strategy *viz.* LinkFailure-FT is similar to the approach adopted by Fortz and Thoroup [9] with some modifications. Another new strategy *viz.* LinkFailure-SS is proposed, where the weights are first optimized for the Failure state. Keeping these weights fixed, all combinations of weights are tried for the added link to find the best cost for the Normal state. Both strategies are discussed further in this paper.

Similar problem has been attempted by Fortz and Thoroup [9]. In their approach, a set of links was considered as critical, and in each iteration one of these links was failed based on the maximum utilization among critical links. The cost of normal topology and the resulting failed topology was averaged and the search was driven to find a solution which minimizes the average cost. One of our implementations in this work is similar to this approach but with the modification that the link failed is always the one connected between nodes carrying the highest demand. This simulates the worst case scenario.

The rest of the paper is organized as follows; The OSPFWS problem statement and the cost functions proposed in the literature are presented in Section 2. The two Link Failure algorithms are discussed and analyzed in Section 3. This is followed by the experimental results including the comparison of both algorithms under Normal and Failure state in Section 4. Finally, we conclude in Section 5.

## 2. PROBLEM STATEMENT AND COST FUNCTION

The OSPF weight setting problem can be stated as follows: Given a directed network of nodes and arcs $G = (N, A)$, a demand matrix $D$, and capacity $C_a$ for each arc $a \in A$, determine a positive integer weight $w_a \in [1, w_{max}]$ for each arc $a \in A$ such that the objective function or cost function $\Phi$ is minimized. When routing is done using OSPF the assigned link weights completely determine the shortest paths, and hence the traffic flows. Based on these traffic flows the partial loads on each arc for a given destination are computed. This is done for all destination nodes. The aggregated partial loads for all destinations on a particular arc give the total load $l_a$ on that arc. The cost of sending traffic through this arc is given by $\Phi_a(l_a)$. The cost value depends on the utilization of the arc and is given by the linear function proposed by Fortz and Thoroup.

$$\Phi'_a(l) = \begin{cases} 1 & for\, 0 \leq l/c_a < 1/3, \\ 3 & for\, 1/3 \leq l/c_a < 2/3, \\ 10 & for\, 2/3 \leq l/c_a < 9/10, \\ 70 & for\, 9/10 \leq l/c_a < 1, \\ 500 & for\, 1 \leq l/c_a < 11/10, \\ 5000 & for\, 11/10 \leq l/c_a < infinity \end{cases} \quad (1)$$

The Fortz cost function is given in equation 2.

$$\Phi = \sum_{a \in A} \Phi_a(l_a) \quad (2)$$

The objective is to minimize $\Phi$, subject to these constraints:

$$l_a = \sum_{(s,t) \in NXN} f_a^{(s,t)} \; a \in A, \quad (3)$$

$$f_a^{(s,t)} \geq 0 \quad (4)$$

In constraint 3, for traffic between source destination pair (s,t), $f_a^{(s,t)}$ indicates the amount of traffic flow that goes over arc a.

The detailed steps showing the formulation of this cost function can be found in the literature [11, 10].

## 3. LINK FAILURE

Handling link failure scenarios requires dealing with two states of a network. In the Normal state, the topology is said to have $n+1$ links. There exist a set of weights W which optimize the cost for this topology. The cost function for this is denoted by $\Phi(n+1)_{OH}$, where OH stands for Original Heuristic and $n+1$ indicates a topology with $n+1$ links. In the case of failure, these weights for the new topology will result in another cost and is denoted by $\Phi([n+1]-1)_{OH}$. Here, $[n+1]-1$ indicates failure of link and topology change from $n+1$ to $n$ links. The above functions are representative of the costs when the Normal state topology was optimized using the original heuristic.

### 3.1. LinkFailure – FT

In LinkFailure – FT strategy, to find optimum weights representing both the normal and the failed states, the idea is not to minimize the cost of each state individually but to minimize the combined or average cost of both states.

For a given solution or set of weights W for the Normal state, the cost is denoted by $\Phi(n+1)$ and for the Failure state with the same set of weights minus the failed link (W-a) the cost is $\Phi`(n)$. The objective is to find the set of weights which minimizes the new cost function:

$$\Phi_{Avg} = \frac{1}{2} (\Phi(n+1) + \Phi`(n)) \quad (5)$$

Starting with a random initial solution for the Normal state ($T_{Norm}$) the same set of weights, except the weight of the failed link, are transferred to the failed state ($T_{Fail}$) and both topologies find the shortest paths and the cost of the initial solution. Tabu Search is started on $T_{Norm}$ by making random moves, and every time the same move is again transferred to $T_{Fail}$.

Both topologies find the shortest path and the corresponding cost after a move. The cost of the new solution for $T_{Norm}$ is denoted as $\Phi(n+1)_{Avg}$ and the new cost of $T_{Fail}$ is denoted as $\Phi(n)_{Avg}$. The cost of the current solution $\Phi_{Avg}$ is the average of $\Phi(n+1)_{Avg}$ and $\Phi(n)_{Avg}$.

$$\Phi_{Avg} = \frac{1}{2} (\Phi(n+1)_{Avg} + \Phi(n)_{Avg}) \quad (6)$$

Here, $\Phi(n+1)_{Avg}$ and $\Phi(n)_{Avg}$ indicate the cost of $T_{Norm}$ and $T_{Fail}$ respectively while optimizing the average cost function. We continue Tabu Search and compute the average cost for each iteration until the termination criteria is met. The set of weights which gives the least value of $\Phi_{Avg}$ is the best solution obtained by the new heuristic. Figure 1 shows the structure of the LinkFailure – FT algorithm.

### 3.2. Performance Evaluation of LinkFailure – FT

The performance of this strategy can be evaluated by comparing the cost obtained for $T_{Norm}$ and $T_{Fail}$ using this heuristic with that of the original. The difference between the costs of the original and the new heuristic would indicate a gain or loss in the solution quality. For $T_{Norm}$, this difference would be:

$$\delta_{Norm} = \Phi(n+1)_{OH} - \Phi(n+1)_{Avg} \quad (7)$$

Optimizing weights using the original heuristic is expected to give a better cost than optimizing for average cost. Hence, the value of $\delta_{Norm}$ is expected to be negative, indicating a loss in solution quality in the Normal state. A smaller $\delta_{Norm}$ value or a value close to zero would indicate that the heuristic is performing well in the Normal state.

In the case of the Failure state the cost difference would be indicated as:

$$\delta_{Fail} = \Phi([n+1]-1)_{OH} - \Phi(n)_{Avg} \quad (8)$$

The purpose of optimizing the weights for link failure is to achieve a better cost in case of a Failure state than would have been achieved with the original heuristic. Hence, $\delta_{Fail}$ must be a positive value indicating an improvement in the solution quality. A larger $\delta_{Fail}$ value would indicate that the new heuristic is performing well in the case of a Failure state. Hence, a combination of smaller $\delta_{Norm}$ value and larger $\delta_{Fail}$ value would be an ideal case indicating minimal loss in the case of the Normal state and significant improvement in the case of the Failure state.

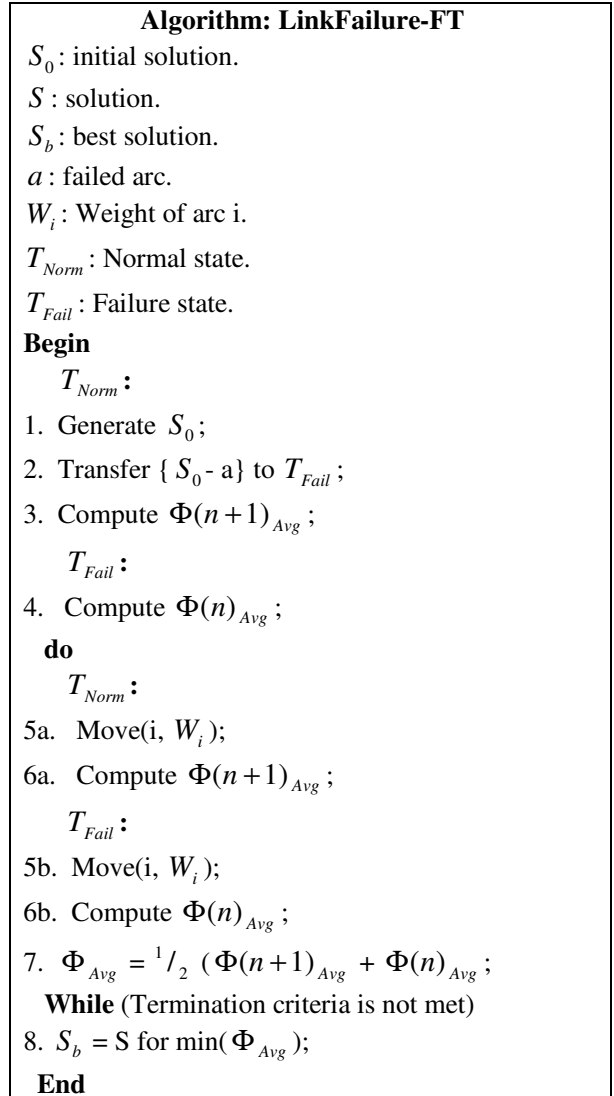

**Algorithm: LinkFailure-FT**
$S_0$: initial solution.
$S$: solution.
$S_b$: best solution.
$a$: failed arc.
$W_i$: Weight of arc i.
$T_{Norm}$: Normal state.
$T_{Fail}$: Failure state.
**Begin**
  $T_{Norm}$:
1. Generate $S_0$;
2. Transfer $\{S_0 - a\}$ to $T_{Fail}$;
3. Compute $\Phi(n+1)_{Avg}$;
  $T_{Fail}$:
4. Compute $\Phi(n)_{Avg}$;
  **do**
  $T_{Norm}$:
5a. Move(i, $W_i$);
6a. Compute $\Phi(n+1)_{Avg}$;
  $T_{Fail}$:
5b. Move(i, $W_i$);
6b. Compute $\Phi(n)_{Avg}$;
7. $\Phi_{Avg} = {}^1/_2 (\Phi(n+1)_{Avg} + \Phi(n)_{Avg}$;
  **While** (Termination criteria is not met)
8. $S_b = S$ for min($\Phi_{Avg}$);
**End**

Figure 1: Structure of the LinkFailure – FT algorithm.

### 3.3. LinkFailure – SS

In the previous strategy, we have tried to optimize weights for the average cost of $T_{Norm}$ and $T_{Fail}$. In this section, we propose another strategy which optimizes weights for $T_{Fail}$ and finds the best solution for $T_{Norm}$ by keeping the weights obtained from $T_{Fail}$ unchanged and trying all possible weights for the one additional link. The test cases and benchmark topologies used were the same as for the previous strategy.

We start with a random initial solution for $T_{Fail}$ and find the shortest paths and corresponding cost for this solution. Tabu Search is started on $T_{Fail}$ by making random moves and after each move, the shortest paths and corresponding cost are computed. The cost of the new solution for $T_{Fail}$ is denoted as $\Phi(n)_{OH}$ which indicates that the cost is for the topology with $n$ links optimized using the original heuristic. Once the termination criterion is met, we obtain the best solution for $T_{Fail}$ and compute its best cost.

The final $n$ weights are transferred to $T_{Norm}$. The weight on the additional $(n+1)^{th}$ link is assigned values from 1 to 20. For each weight $W_i$, the cost of the $i^{th}$ solution is computed. The twenty costs obtained are compared to find the best solution for $T_{Norm}$. This is denoted by $\Phi([n]_{OH}+1)_{20}$ which indicates that the cost is for topology with $n+1$ links where $n$ links are optimized using the original heuristic and one additional link is optimized by finding the best solution from the twenty possible combinations. Figure 2 shows the structure of the LinkFailure – SS algorithm

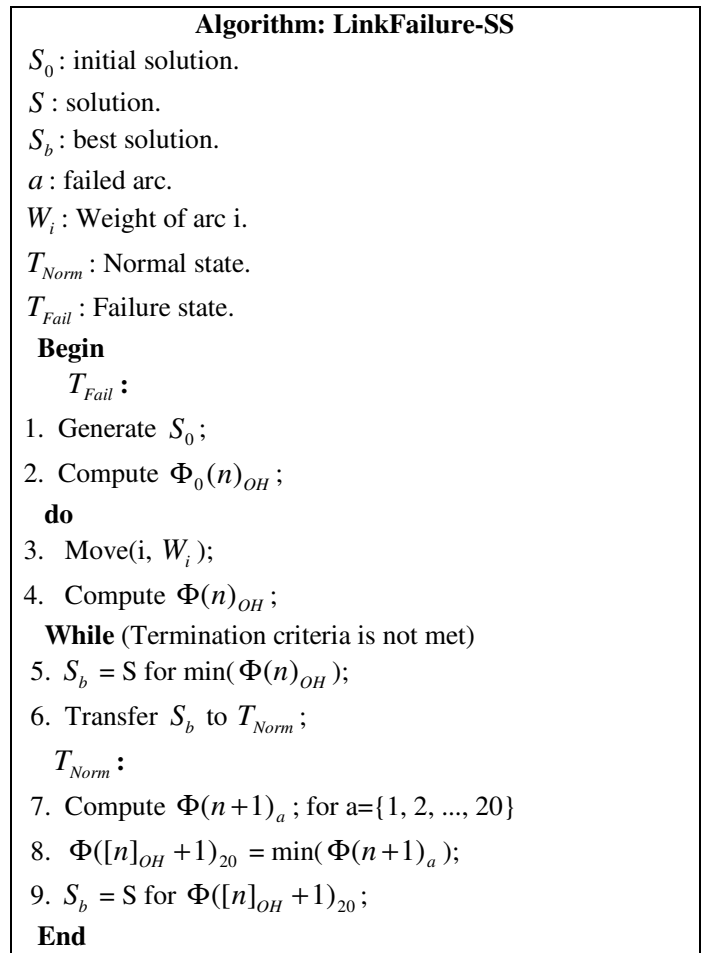

**Algorithm: LinkFailure-SS**

$S_0$: initial solution.
$S$ : solution.
$S_b$: best solution.
$a$ : failed arc.
$W_i$ : Weight of arc i.
$T_{Norm}$ : Normal state.
$T_{Fail}$ : Failure state.
 **Begin**
    $T_{Fail}$ :
1. Generate $S_0$;
2. Compute $\Phi_0(n)_{OH}$;
  **do**
3. Move(i, $W_i$);
4. Compute $\Phi(n)_{OH}$;
   **While** (Termination criteria is not met)
5. $S_b$ = S for min($\Phi(n)_{OH}$);
6. Transfer $S_b$ to $T_{Norm}$;
   $T_{Norm}$ :
7. Compute $\Phi(n+1)_a$; for a={1, 2, ..., 20}
8. $\Phi([n]_{OH}+1)_{20} = \min(\Phi(n+1)_a)$;
9. $S_b$ = S for $\Phi([n]_{OH}+1)_{20}$;
 **End**

Figure 2: Structure of the LinkFailure-SS algorithm.

International Journal of Computer Networks & Communications (IJCNC) Vol.3, No.1, January 2011

### 3.4. Performance Evaluation of LinkFailure – SS

Similar to the FT approach, the performance of this strategy can be evaluated by comparing the cost obtained for $T_{Norm}$ and $T_{Fail}$ using the SS heuristic with that of the original heuristic (OH). The difference between costs of the original and the new heuristic would indicate a gain or loss in the solution quality. For $T_{Norm}$ this difference would be:

$$\delta_{Norm} = \Phi(n+1)_{OH} - \Phi([n]_{OH} +1)_{20} \quad (9)$$

In the case of the Failure state, the original heuristic will end up with a cost $\Phi([n+1]-1)_{OH}$ and the SS heuristic with a cost of $\Phi(n)_{OH}$. Hence, the cost difference would be indicated as:

$$\delta_{Fail} = \Phi([n+1]-1)_{OH} - \Phi(n)_{OH} \quad (10)$$

In the SS approach, the weights are optimized for $T_{Fail}$ and are expected to achieve a better cost in the case of a link failure than would have been achieved with the original heuristic. Hence, $\delta_{Fail}$ must be a positive value indicating an improvement in the solution quality.

### 3.5. FT versus SS

In the case of LinkFailure-FT, we simultaneously optimize two states of a network *viz.* $T_{Norm}$ and $T_{Fail}$, whereas in LinkFailure-SS we only optimize $T_{Fail}$ and then try the best possible weight for the one additional link to optimize $T_{Norm}$. Hence, the SS approach has a faster convergence when compared to FT; which is a major factor when dealing with larger networks and higher demands.

As discussed earlier, SS is optimized for the Failure state and hence should not only give better solution when compared to OH but also should perform better than FT in the Failure state. In the FT approach, the weights are selected to optimize the average cost and not the best cost for individual states. Any heuristic, to be acceptable, must not degrade the performance of the network in the Normal state. In other words it should result in a solution quality as close to the Original Heuristic (OH) as possible.

## 4. RESULTS

In this section, we present the experimental results for the two heuristics mentioned in the previous section. The benchmarks used for the evaluation of the original heuristic for no failure case [12, 11] were also used for the link failure case. Due to the change in topology (different number of links) in the two states, the original test case would represent only one of the states and a modified test case would represent the other state.

Representing the failed state with a modified test case would require deletion of the corresponding link entries from the files representing the graph and capacity of links. This could also result in a disconnection of the graph. To avoid this, we represented the Failure state with the original test cases. To represent the Normal state, we add an additional link between two nodes $n1$ and $n2$. The nodes selected were the ones with the highest demand between them in the demand matrix. Failing this particular link which is directly connected between the two nodes having the highest demand between them would cause the worst effect on the network.

174

Hence, if our heuristic is able to optimize weights for the worst case scenario then it is expected to be robust.

The notations used to denote Cost in the Normal and Failure state are shown below:

| $FT_N$ | $SS_N$ | $OH_N$ |
|---|---|---|
| $\Phi(N+1)_{avg}$ | $\Phi([N]+1)_{20}$ | $\Phi(N+1)_{oh}$ |

| $FT_F$ | $SS_F$ | $OH_F$ |
|---|---|---|
| $\Phi(N)_{avg}$ | $\Phi(N)_{oh}$ | $\Phi([N+1]-1)_{oh}$ |

### 4.1. FT versus OH

Experimental results for the two strategies implemented for the single link failure scenario are presented in this section. The individual performance of each strategy can be evaluated by comparing its results in Normal and Failure States to the Original Heuristic.

Table 1 shows the Cost values obtained using FT Strategy and OH for five different demands using the test case h100N360a. From the table, it can be seen that in the Normal state the Cost of FT is marginally higher than OH, which can be seen in the $\delta_{Norm}$ column which shows the Cost difference for the two strategies in the Normal State. Negative values indicate a loss. As expected, there is some loss in the Normal State. In the Failure State, for all demands except Demand-9, the FT Cost is less than the OH Cost as indicated by a positive value in the column $\delta_{Fail}$. Hence, there is some gain in the Failure State. The overall gain or loss is indicated in the column $\delta$.

The value of $\delta_{Norm}$ is more than the value of $\delta_{Fail}$ for higher demands D11, D12 which implies that the margin of loss in Normal state is more than the gain in the Failure State for this case at higher demands. Results also show an overall gain for the two demands D8 and D10.

Table 1: Cost Comparison FT versus OH in Normal and Failure State for h100N360a Network.

| D | $FT_N$ | $OH_N$ | $\delta_N$ | $FT_F$ | $OH_F$ | $\delta_F$ | $\delta$ |
|---|---|---|---|---|---|---|---|
| D8 | 1.313 | 1.320 | 0.006 | 1.336 | 2.743 | 1.406 | 1.413 |
| D9 | 1.482 | 1.448 | -0.033 | 1.538 | 1.494 | -0.044 | -0.077 |
| D10 | 2.096 | 1.985 | -0.111 | 2.315 | 5.711 | 3.396 | 3.285 |
| D11 | 4.498 | 4.369 | -0.129 | 6.017 | 6.057 | 0.040 | -0.089 |
| D12 | 17.973 | 14.076 | -3.897 | 24.398 | 25.487 | 1.089 | -2.809 |

## 4.2. SS versus OH

Table 2 shows similar comparison for the SS Strategy. Even in this case, the values of $\delta_{Norm}$ for SS are marginally higher than those of OH, and the values of $\delta_{Fail}$ for SS are well below those of OH for all five demands shown in the table. This shows that there is a slight loss in the Normal State and a significant improvement in the Failure State. There is an overall gain as indicated by a positive values in the last column $\delta$. Hence, there is an improvement in performance due to the use of SS strategy compared to OH.

Table 2: Cost Comparison SS versus OH in Normal and Failure State for h100N360a Network.

| D | $SS_N$ | $OH_N$ | $\delta_N$ | $SS_F$ | $OH_F$ | $\delta_F$ | $\delta$ |
|---|---|---|---|---|---|---|---|
| D8 | 1.329 | 1.320 | -0.009 | 1.343 | 2.743 | 1.399 | 1.390 |
| D9 | 1.480 | 1.448 | -0.032 | 1.487 | 1.494 | 0.007 | -0.025 |
| D10 | 1.986 | 1.985 | -0.001 | 2.010 | 5.711 | 3.701 | 3.700 |
| D11 | 4.389 | 4.369 | -0.019 | 5.330 | 6.057 | 0.727 | 0.708 |
| D12 | 14.316 | 14.076 | -0.240 | 18.158 | 25.487 | 7.329 | 7.089 |

## 4.3. FT versus SS

We have seen that both strategies are performing better than the Original Heuristic in the Failure state while OH has slightly better results for the Normal state. We now compare the SS and FT results to show which of the two heuristics performs better. The comparison is shown in Table 3.

In the Normal state, for the demands D8 - D10 both strategies have almost the same cost values with marginal differences in favour of SS. For the highest demand D12, SS clearly performs better than FT. Overall, for the Normal State, it can be said that SS performs better than FT for this test case. For the Failure State, SS clearly outperforms FT for all demands. This is expected as the strategy is specifically designed to optimize weights for the Failure State or in other words to minimize the Failure State Cost. Hence, SS is always expected to produce better results for a Failure State. The overall comparison shows superiority of SS over FT for this test case. Comparison of all three strategies for this test case is presented below.

Table 3: Cost Comparison FT versus SS in Normal and Failure State for h100N360a Network.

| Demand | $FT_N$ | $SS_N$ | $FT_F$ | $SS_F$ |
|---|---|---|---|---|
| D8 | 1.31326 | 1.32905 | 1.33621 | 1.34312 |
| D9 | 1.48152 | 1.48005 | 1.53819 | 1.48682 |
| D10 | 2.09604 | 1.98619 | 2.31527 | 2.01001 |
| D11 | 4.49806 | 4.38878 | 6.01734 | 5.33037 |
| D12 | 17.9732 | 14.3157 | 24.3984 | 18.1582 |

## 4.4. OH versus FT versus SS

Figure 3 shows the graph with the Cost comparison of all the three heuristics in the Normal state and in Figure 4 for the Failure State for the h100N360a Network.

In Figure 3, it can be seen that OH has the best Cost in the Normal state which is very closely matched by SS. FT comparatively has the worst Cost in the Normal state. In the Failure state SS outperforms both FT and OH as seen in Figure 4. Hence, SS has proved to be having a marginal loss (negligible in the case of lower demands) in the Normal state and a significant gain in the case of Failure, which is the ideal requirement for these types of problems.

Experiments were conducted for five more test cases and are presented in the following figures for both the normal and failure states. All these figures provide a comparison for all three algorithms, i.e., FT, SS, and OH. A summary of the results obtained is presented at the end of this section.

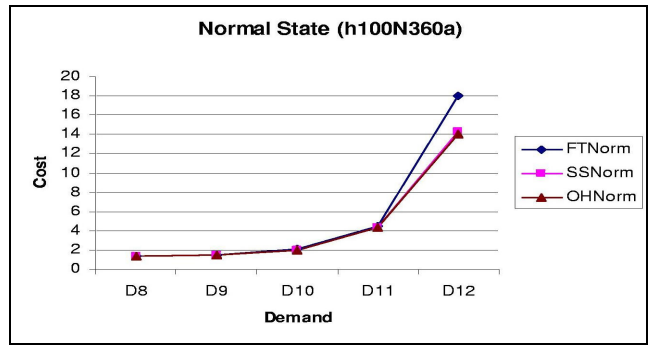

Figure 3: Cost Comparison FT, SS and OH in the Normal state for h100N360a Network.

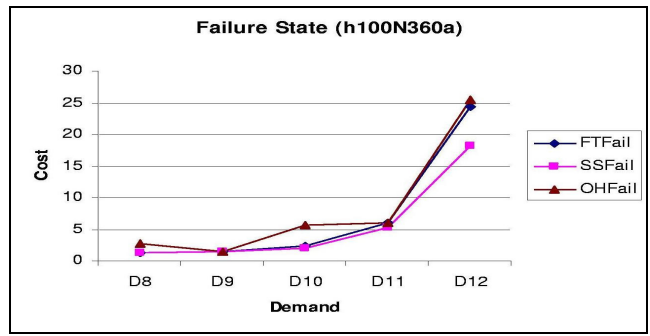

Figure 4: Cost Comparison FT, SS and OH in the Failure state for h100N360a Network.

Figure 5 shows the graph with the Cost comparison of all the three heuristics in the Normal state and in Figure 6 for the Failure State for the r50N228a Network.

In Figure 5, it can be seen that both SS and FT show comparable results in the Normal state. In the Failure state, SS outperforms both FT and OH as seen in Figure 6.

International Journal of Computer Networks & Communications (IJCNC) Vol.3, No.1, January 2011

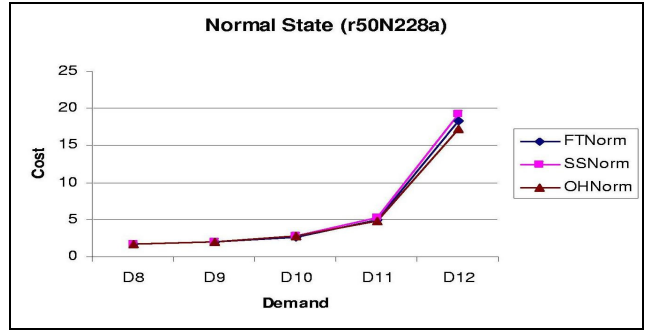

Figure 5: Comparison of FT, SS, and OH in the Normal state for r50N228a Network.

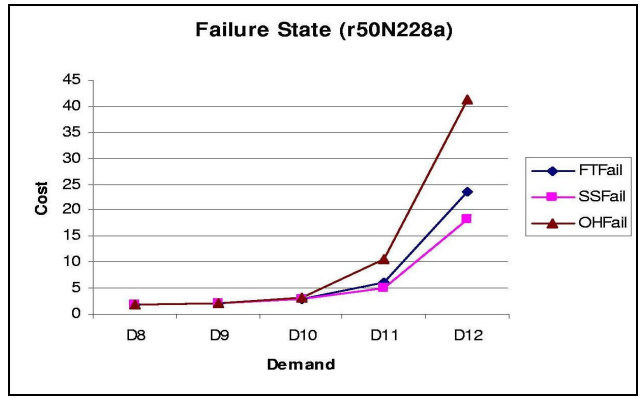

Figure 6: Comparison of FT, SS, and OH in the Failure state for r50N228a Network.

Figure 7 shows the graph with the Cost comparison of all the three heuristics in the Normal state and in Figure 8 for the Failure State for the r100N503a Network.

In Figure 7, it can be seen that both SS and FT show comparable results in the Normal state. In the Failure state, SS outperforms both FT and OH as seen in Figure 8.

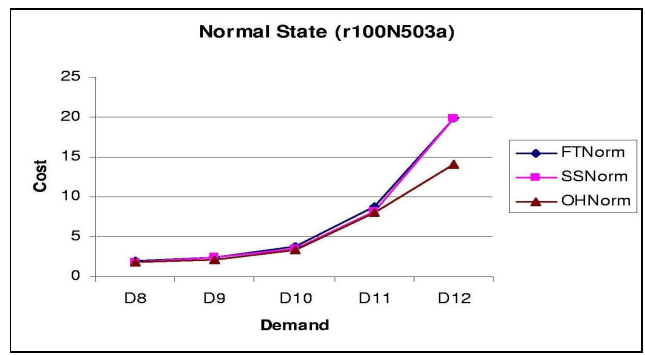

Figure 7: Comparison of FT, SS, and OH in the Normal state for r100N503a Network.

178

International Journal of Computer Networks & Communications (IJCNC) Vol.3, No.1, January 2011

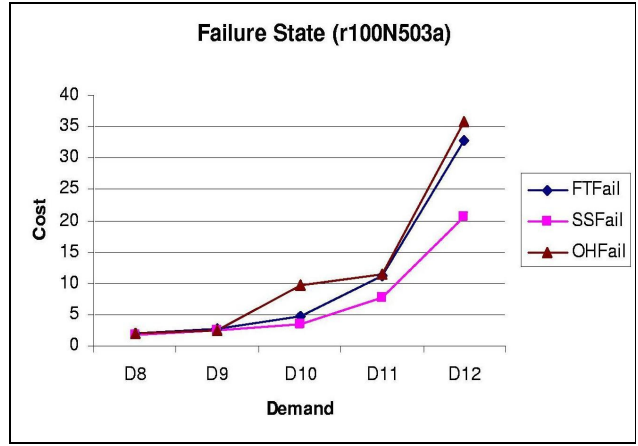

Figure 8: Comparison of FT, SS, and OH in the Failure state for r100N503a Network.

Figure 9 shows the graph with the Cost comparison of all the three heuristics in the Normal state and in Figure 10 for the Failure State for the w50N169a Network.

In Figure 9, it can be seen that all strategies perform equally well in the Normal state for all demands. In the Failure state, similarly all strategies perform equally well for all demands as seen in Figure 10. This indicates that a link failure does not have significant effect on network performance for this test case.

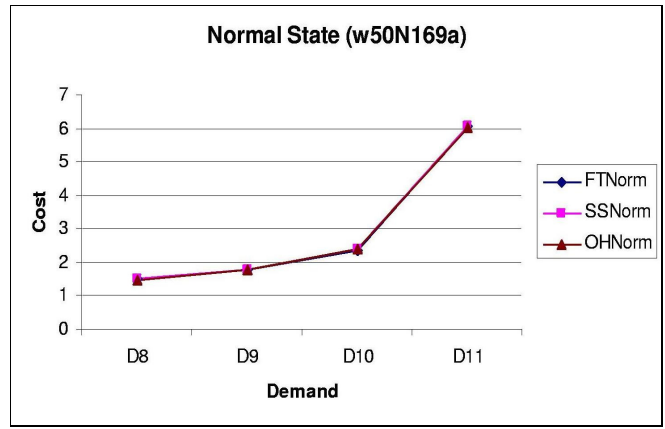

Figure 9: Comparison of FT, SS, and OH in the Normal state for w50N169a Network.

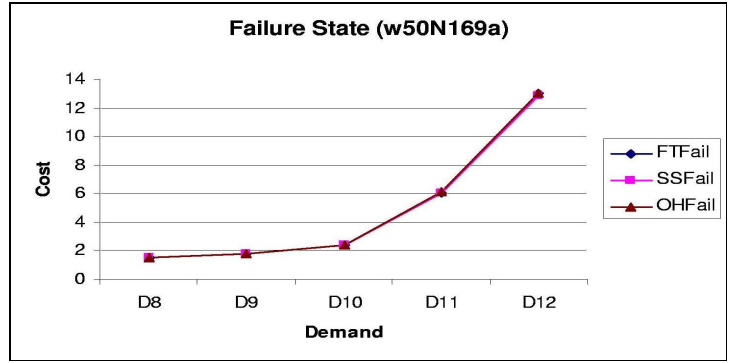

Figure 10: Comparison of FT, SS, and OH in the Failure state for w50N169a Network.

Figure 11 shows the graph with the Cost comparison of all the three heuristics in the Normal state and in Figure 12 for the Failure State for the w100N476a Network.

In Figure 11, it can be seen that all strategies perform equally well in the Normal state for all demands. In the Failure state, similarly all strategies perform equally well for all demands as seen in Figure 12. This indicates that a link failure does not have significant effect on network performance for this test case.

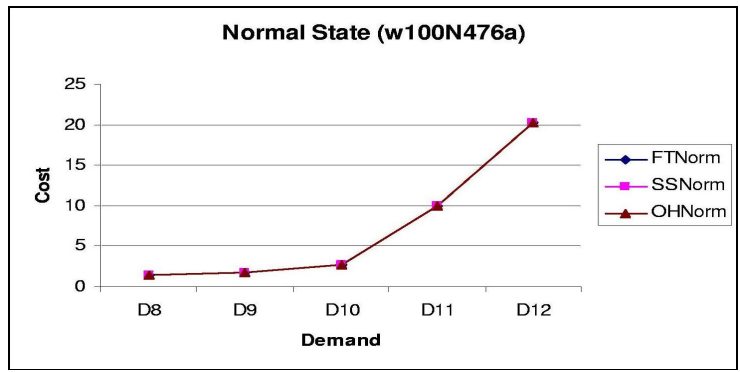

Figure 11: Comparison of FT, SS, and OH in the Normal state for w100N476a Network.

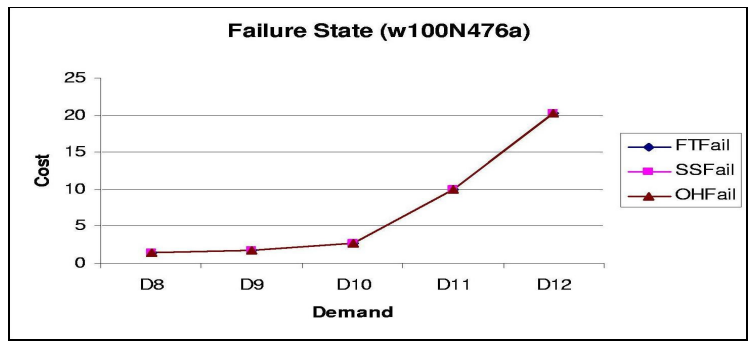

Figure 12: Comparison of FT, SS, and OH in the Failure state for w100N476a Network.

Figure 13 shows the graph with the Cost comparison of all the three heuristics in the Normal state and in Figure 14 for the Failure State for the h50N148a Network.

In Figure 13, it can be seen that OH has the best Cost in the Normal state which is very closely matched by SS. FT comparatively has the worst Cost in the Normal state. In the Failure state SS outperforms both FT and OH as seen in Figure 14.

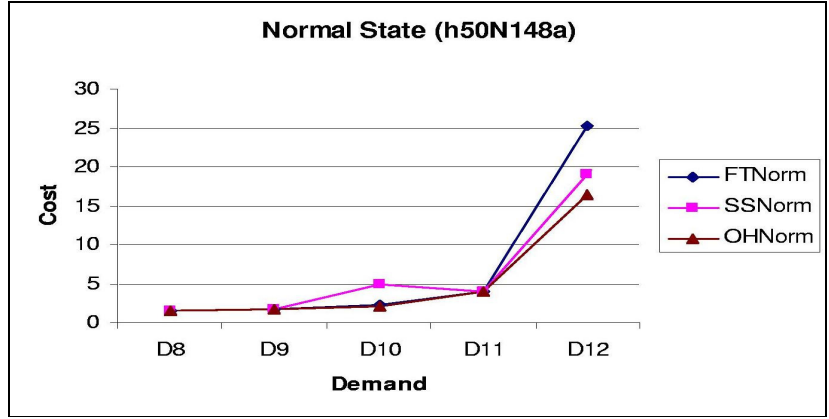

Figure 13: Comparison of FT, SS, and OH in the Normal state for h50N148a Network.

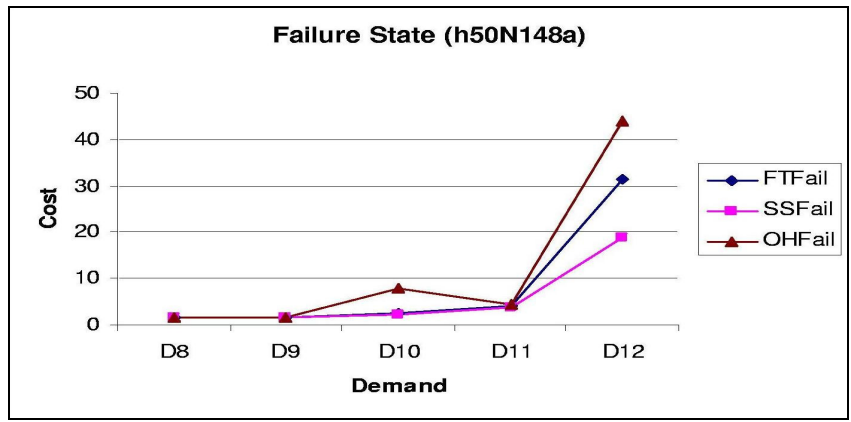

Figure 14: Comparison of FT, SS, and OH in the Failure state for h50N148a Network.

### 4.5. Summary of Results

In all the test cases, SS achieves the best results for the Failure state ($\delta_{Fail}$) and also for the overall improvement ($\delta$). SS is followed by FT in the Failure state, which performs better than OH. In the Normal state, SS performs slightly better than FT for the two test cases h50N148a and h100N360a and has comparable results for the two cases r100N503a and r50N228a. For the two Waxman graphs, w50N169a and w100N476a, all strategies perform equally well in Normal and Failure state for all demands. This indicates that a link failure does not have significant effect on network performance for these two cases. Finally, it can also be observed that for lower demands (Demand-8, Demand-9), the results are almost the same for all the six test cases. This indicates that, if the load on the network is low, there is minimum effect of the link failure on the network performance and the original heuristic itself is efficient enough to handle single link failures.

## 5. CONCLUSIONS

The single link failure issue in OSPF routing was addressed in this work to find a weight setting for the links which results in efficient routing in normal and failure states. Two new heuristics based on Tabu Search were proposed in this paper, namely LinkFailure-SS and LinkFailure-FT. Both heuristics were evaluated and they both produced better results when compared to the original heuristic in the Failure state. In addition, the SS approach is found to give better results than the FT approach in both normal and failure states. Therefore, it can be concluded that the SS approach is an efficient way to tackle single link failure issues. It was also shown through experimental results that at lower demands and traffic loads the effect of link failure on network performance is less and the original heuristic can also handle single link failures if the traffic load on the network is low.

## ACKNOWLEDGEMENTS

Acknowledgement goes to KFUPM for supporting this research work. This material is based in part on work supported by a KFUPM project under Grant No. SAB-2006-10. The authors wish to thank Bernard Fortz and Mikkel Thorup for sharing the test problems.

International Journal of Computer Networks & Communications (IJCNC) Vol.3, No.1, January 2011

[12]  W. Zegura (1996) GT-ITM: Georgia Tech internetwork topology models (software), *http://www.cc.gatech.edu/faq/Ellen.Zegura/gt-itm/gt-itm.tar.gz*.

**Authors**

Mohammed H. Sqalli received a degree of "Ingenieur d'Etat" in Computer Science from Ecole Mohammadia d'Ingenieurs, Rabat, Morocco in 1992. He earned a Master's degree in Computer Science in 1996 and a Ph.D. degree in Engineering - Systems Design in 2002, both from the University of New Hampshire, Durham, NH, USA. He is a recipient of a Fulbright Scholarship for the period of 1994-1998. Mohammed H. Sqalli is currently an assistant professor in the Computer Engineering Department at KFUPM. He is also an IEEE member. His research interests include: Network Security, Cloud Computing, Network Design and Management, Traffic Engineering, and Iterative Heuristics. He has over 25 publications in related areas.

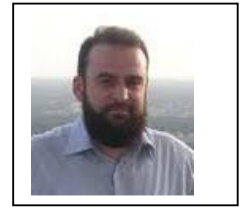

Sadiq M. Sait obtained a Bachelor's degree in Electronics from Bangalore University in 1981, and Master's and PhD degrees in Electrical Engineering from King Fahd University of Petroleum & Minerals (KFUPM), Dhahran, Saudi Arabia in 1983 & 1987 respectively. Sadiq M. Sait is the co-author of the book VLSI PHYSICAL DESIGN AUTOMATION: Theory & Practice, published by McGraw-Hill Book Co., Europe, (and also co-published by IEEE Press), January 1995, and ITERATIVE COMPUTER ALGORITHMS with APPLICATIONS in ENGINEERING (Solving Combinatorial Optimization Problems): published by IEEE Computer Society Press, California, USA, 1999. He was the Chairman of Computer Engineering Department, KFUPM from January 2001 - December 2004. Presently he is the Director of Information Technology Center (ITC) at KFUPM, since January 2005.

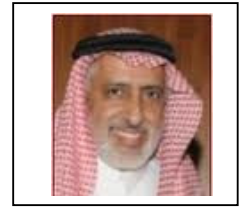

Syed Asadullah received a Bachelor of Technology (B. Tech) degree in Electronics and Communications Engineering from Jawaherlal Nehru Technological University, Hyderabad, India in 2000. He also obtained his Master's degree in Computer Networks from KFUPM, Saudi Arabia in 2008.

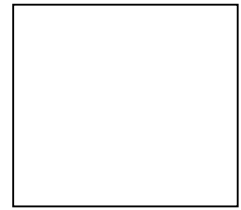